# Exploring Invariants & Patterns in Human Commute time

Hongyan Cui [1, 2, #], Yuxiao Wu [1, *], Stanislav Sobolevsky [2,3, &], Shuai Xu [1,@], Carlo Ratti [2, ^]

1. State Key Lab of Networking and Switching Technology, Key Lab. of network system architecture and convergence，Beijing University of Posts and Telecommunications, China.

2. Massachusetts Institute of Technology, 77 Massachusetts Avenue, Cambridge, 02139, USA.

3. New York University，1 MetroTech Center, Office 1910, Brooklyn, NY

# cuihy@bupt.edu.cn (corresponding author)

* wuyuxiao@bupt.edu.cn

& stanly@mit.edu

@ awsxsa@hotmail.com

^ ratti@mit.edu

**Abstract**

In everyday life, the process of commuting to work from home happens every now and then. And the research of commute characteristics is useful for urban function planning. For humans, the commute of an individual seems revealing no regular universal patterns, but it is true that people try to find a satisfactory state of life regarding commute issues. Commute time and distance are most important indicators to measure the degree of this satisfaction. Marchetti states a certain regularity in human commute time distribution – specifically, it states that no matter when, where and how far away people live, they always tend to spend approximately the same average time for their daily commute. However, will the rapid development of cities nowadays as well as serious challenges brought by economic development affect this constant? If there are novel characteristics? We revisit these problems using fine grained communication data in two Chinese major cities during recent two years. The results indicate that the commute time has been slightly increased from Marchetti's constant with the development of society. People's overall travel budgets have been increased, more concretely speaking, for medium and long distance commuters, their endurance limit for commute time is enhanced during the passing years, and fluctuates around a constant; for short distance commuters, their commute time increases with the distance. Moreover, the population distribution in every commute distance shows strong cross-city similarity and does not change much over two years.

**Introduction**

**Human mobility**

Researchers use various datasets to explore the human mobility, such as detailed census data[1], mobile phone call detail records(CDR)[ 2, 3, ,4, 5, 6, 7], taxi trip records [8, 9, 10], Tweets [11], Flickr[12], Foursquare check-in data [13], GPS devices [14, 15, 16] and even circulating bank notes [17, 18]. The human behavior researches have far-reaching implications in communication network planning, urban planning, and infrastructure construction. Kang et al. integrates mobile phone and taxicab usages together to explore human movements in Singapore, and uncovers substantial differences between taxi cab trips and mobile phone movements in terms of spatial distribution [10]. They confirm that for long distance commuters, the users' amount of taxicab trips decays with distance more quickly compared to mobile phone users' amount. Therefore, it's beneficial to use a mobile phone dataset to get more accurate result. Hawelka et al. apply almost a billion Tweets records in 2012 to estimate the volumes of international travelers [11]. They find that the temporal patterns disclose the universal seasons of increased international mobility and the peculiar national nature of overseen travels. Noulas et al. use the Foursquare data to propose that the variations of human movements in different metropolis are mainly due to the spatial distribution of places in the city, other than other potential factors such as cultural cognition differences [13]. Vazquez et al. use Global Positioning System (GPS) data-loggers to investigate the mobility patterns of 582 residents from two neighborhoods from the city of Iquitos, Peru. They discover that only up to 38% of the tracked participants show a regular and predictable mobility routine, a sharp contrast to the situation in the developed world. Furthermore, they verify that temporally unstructured daily routines increase an epidemic's final spread size [16]. Brockmann et al. report on a solid and quantitative assessment of human travel statistics by analyzing the circulation of the bank notes in the United States, and indicate that the distribution of traveling distances decays as a power law [17].

**Commuting patterns**

The Marchetti's assumption posits that humans, since Neolithic times, budget approximately one hour per day on travel, independent of location, modes of transport, and other lifestyle considerations, it describes the universal uniformity of human commute time [19].

Researchers have reported their researches on this topic, and give out that the daily time constant is 1.1 hours [20], 1.2 hours [21] and 1.3 hours [22]. Yakov Zahavi proposes a similar concept of people's "travel time budget", and he realizes that travelers do not save time as a result of increase on travel speed, but that they apply the time saved from some trips for additional travel. That is to say people seem to have a constant "travel time budget" [23]. David Metz also refers to the travel time budget as a stable daily amount of time that people make available for travel [24]. These work casts doubt on the contention that investment in infrastructure saves travel time. Instead it appears that people invest travel time saved in traveling a longer distance. Kung K. S. et al. use the CDR data to investigate into the constant travel budget hypothesis. They discover that the home-work time are indeed independent with distance in most of the investigated districts, which have multimodal commute behaviors, like Portugal, Ivory Coast, and Boston, for >5km medium/long commutes. They do not conclude for shorter commutes due to limitations on the CDR data, because for short commutes whose trajectories are respectively comparable to the distances between cell towers, their commuting distance would be estimated wrongly; besides, although there was also an attempt to employ cell phone data for research purpose, temporal sparsity of regular CDRs does not allow to draw robust conclusions; while here we introduce a new much more fine-grained dataset providing a principally new level of ground-truth. Another conclusion is that in car-only or car heavy countries, the commute time is seriously influenced by commute distance, like Milan and Saudi Arabia. The commuting distance will influence people's performances in many daily life aspects [25]. Thomas et al. explore the relationship between commuting length and time spent with others, and find that a certain amount of increase on commute time will bring a certain amount of decrease of time spent with families [26]. Besides, Martijn et al. evidence that students with long commute times have lower average grades [27]. Then, what will actually affect the time people spend on commuting? Mcquaid et al. find that the worker's age, having children, the age of their youngest child, occupation, weekly pay, and mode of transport are important factors that affect human commute time. While, their influence cannot be simply described [28]. David analyzes impacts on commuting behavior as a result of workplace relocation in Lisbon, Portugal, showing no insignificant change in commute time, which demonstrates a strategy that aims to maintain commute time within acceptable limits [29].

**Choice about the data sets**

Some researchers point out the effectiveness of datasets for getting the research result recently. [13, 30] point out that temporal–sparsity leads to an inaccuracy of human trajectory, which refers to the insufficiency of sampling. The spatial-sparsity is mainly caused by low resolution of cell towers, too sparse distribution of cell towers would bring non-ignorable inaccuracy on people's positions. Diao et al. try to solve temporal-sparsity problem by combining datasets together, such as mobile trace data, phone calls, short messages records etc. [31]. In this paper, we alleviate the effects of temporal-sparsity and spatial-sparsity by using the subscribers' communication records of one telecommunication operator in two major cities of China (Chongqing and Tianjin). As we all know, cellular network provides people services by frequent connecting the mobile terminal to the Base Station (BS). Many subscribers generate thousands of records when they use various mobile Internet applications and messages-push services, so we can confirm that the data traffic records is enough for depicting people's trajectories during a day, at least more reasonable than simply use call records or Foursquare records.

**Computing the commute distance**

There are various computing methods in literature to estimate the commute distance. For example, the crow-fly distance (either the great circle distance or the Euclidean distance) [32], the shortest distance path [33], the shortest time path [34]. The great circle distance is the distance of the shortest arc distance between two locations on the earth, which would definitely underestimate the real commute distance, and commonly needs a correction factor. Normally speaking, different commuting distance and transportation modes would bring different correction factors, while because of lack of detailed information, we simply regard the great circle distance as commute distance [25]. Due to the relatively high road density in City-A and City-B, and that all results would be affected by estimating in a similar manner, we assume that this potential bias can be limited and ignored.

**Results**

Three communication record datasets are used to investigate the commute patterns, which come from City-A (Chongqing City) in 2012, City-A in 2014 and City-B (Tianjin City) in 2014.

Table 1 gives out the amount of effective users in the test. We filter out unqualified users and leave 8500 active users in city-A (2012), 39,000 in City-A (2014) and 20,600 in City-B (2014). Table 1 reveals the percentages of users with different commuting distance in different years and cities. Regardless of cities and years, about 50%- 57 % of people commute within 2km distance, 23%-25 % of people commute within 2-6km, 14%-17 % of people commute within 6-15km, 3%-4 % of people commute within 15-25km, and only 3%-4 % of people commute more than 25km. This indicates most users prefer to choose their home near their work place.

Fig.1 shows that commuting distance has strong cross-city similarities, since people of different commuting distance almost follow the same distribution in different test cities during two years.

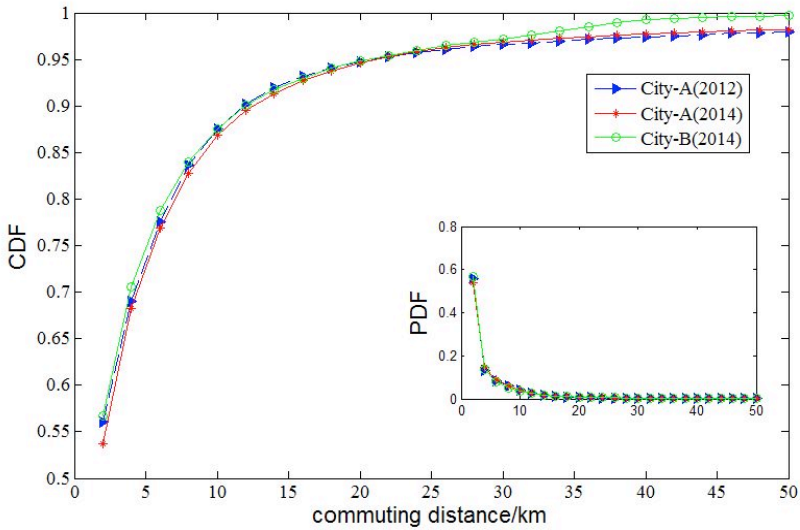

Figure 1. Commute distance distribution

In Fig.2, we find that the people's commute time follows different laws. When commuting distance is less than about 18km, people's commute time simply increases with the commute distance, especially obvious for whose commuting distance is less10km. That is to say, for individual with a commuting distance of less than 18km, it is effective to shorten his commute time by shortening commuting distance, and this may be the reason why many people move their home place constantly closer to their work place; however, when the commuting distance is longer than 18km anyhow, people are no longer keen to pursue the shorter commuting distance. Besides, the increase of approaching constant commute time of long distance commuters during the past two years indicates that people enhance their endurance for maximum commute time as a result of urbanization development.

The laws of commute time summarized from our analyses are presented in Table 3, the subscribers with commute distance of more than 18km show steady travel budget. And for these part of users, their commute constant of single way is 0.80 hour in the morning, and 0.84 hour at night in City-A (2012). However, in 2014, the City-A's morning commute constant and night commute constant respectively rise to 1.34 hours and 1.39 hours. It is similar that in City-B (2014), the morning commute constant is 1.50 hours, and the night commute constant is 1.70 hours. For the users whose commuting distance is less than 18km, their commute time increases with the commute distance.

According to our statistic results, the average commute time of all of the users in the morning and night in City-A (2012) respectively is 0.54 hour (32 minutes) and 0.64 hour (38.4 minutes); that in City-A (2014) is 0.71 hour(42.6 minutes) and 0.88 hour(52.8 minutes) ; and in City-B(2014) it is 0.91 hour (54.6 minutes) and 0.83 hour (49.8 minutes). On average, citizens in City-A spend around more 10.6 minutes in the morning and more 14.4 minutes at night on commuting in 2014 than 2012. We can infer from our statistic results that the average time people spend on transit per day in City-A has increased from 1.18 hour in 2012 to 1.59 hour in 2014 by, and in City-B, this budget in 2014 is 1.74 hour. This suggests that the Marchetti's assumption is slightly increased with the urbanization development. That is to say, the underlying factors that lead to Marchetti's constant still hold true in general, but this constant slightly increases with the development of society and change of times in China.

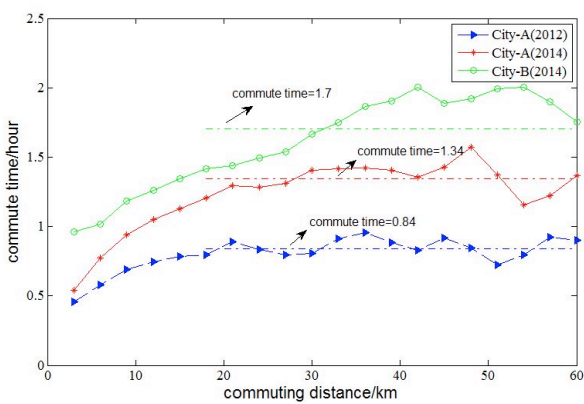

(a) Morning commute time

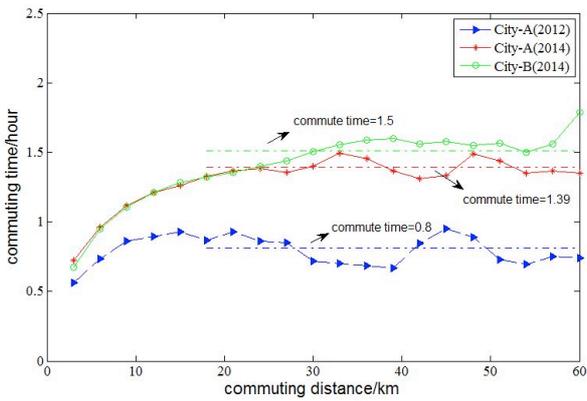

(b) Night commute time

Figure.2. Average commute time under every commute distance. Step size is 3km

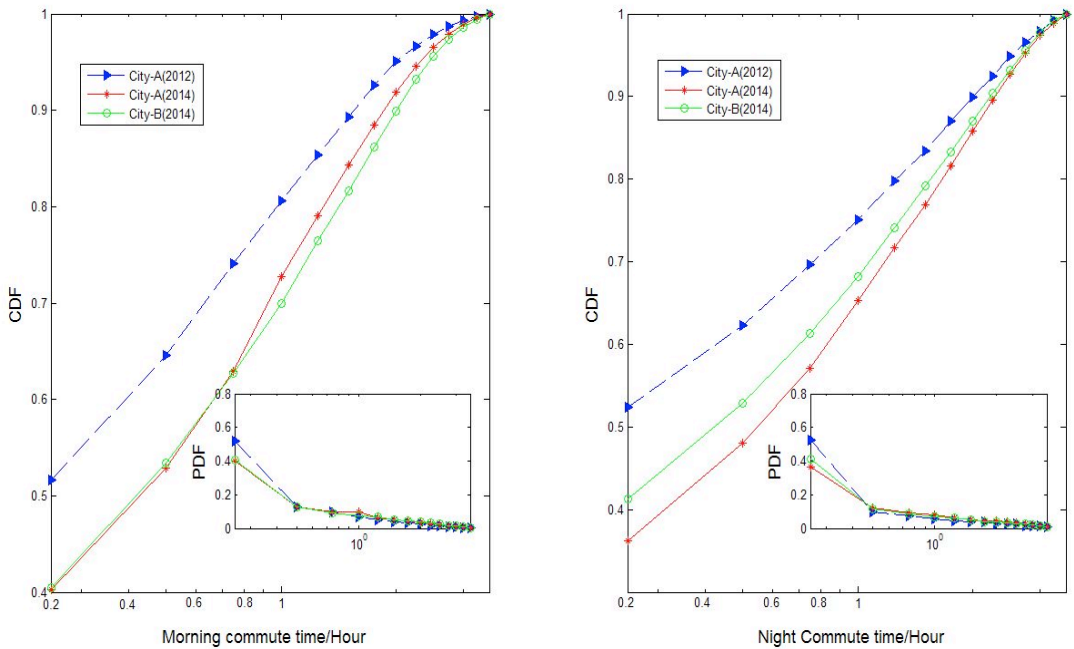

Figure 3. The distributions of users by commute time.

We could get the results in Fig.3 that 80% City-A users' morning commute time in 2012 is less than 1 hour, but this proportion is decreased to 73% in 2014. And in the evening, this proportion also falls from 75% to 65% over the two years, The data indicates that at least 7-10% users increase the commute time within two years in City-A. In City-B, 70% users' morning commute time is less than 1 hour, but for night commute time, this proportion slightly decreases to 67%. It is also easy to discover that people are more likely to commute within 1 hour in the morning than at night regardless of times and city.

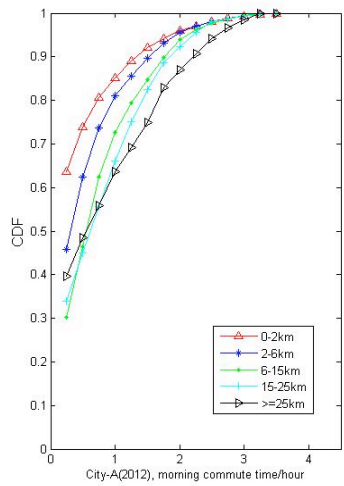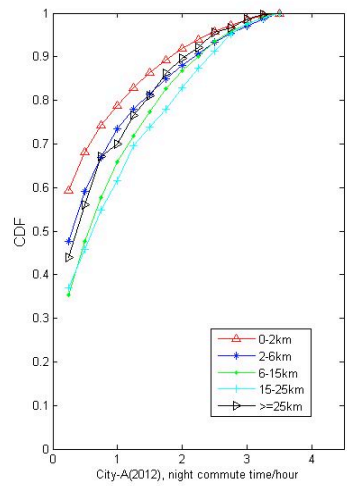
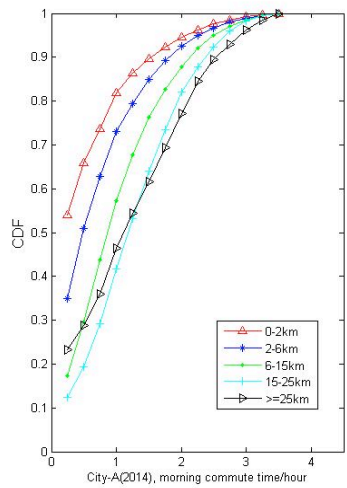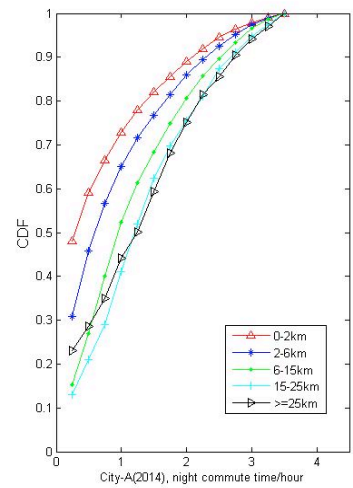

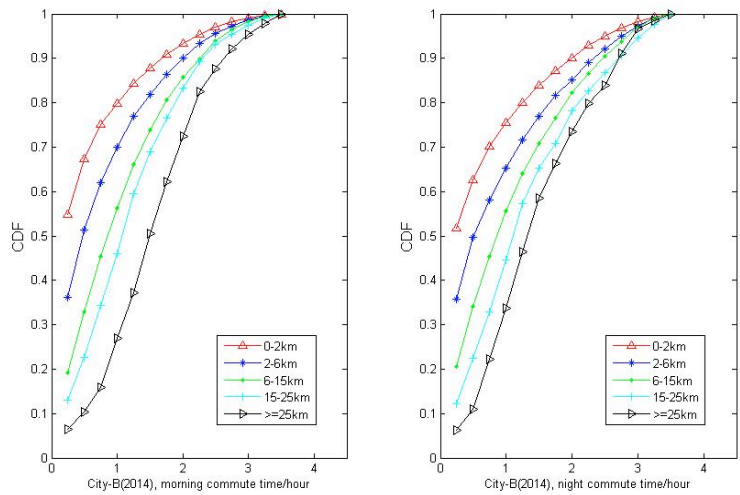

Figure 4. Commute time distributions under every commute distance section in City-A (2012), City-A (2014) and City-B (2014).

In order to investigate commute time patterns among different groups of people, we divide our samples into 5 groups based on different commuting distance, that are 0-2km, 2-6km, 6-15km, 15-25km, and more than 25km distance groups. Each commute distance group is denoted by different colors in Fig. 4.

In Figure 4, it is obvious that in City-A (2012), 64% of short distance commuters (<2km) spend less than 0.25 hour on commuting, and 15% of short distance commuters (<2km) spend one hour or more. There are 33% of long distance users (15-25km) commute less than 0.25 hour, and 45% of long distance commuters (15-25km) spend one hour or more. As a whole, shortening commuting distance would gain commute time for most of people, but for rest of people, using faster transportation vehicles, or adjusting their working periods would help reduce their commute time. For example, for commuters of 2-6km, 50% users spend less than 20 minutes on commuting, this speed is exactly like the speed of riding a bike, so we guess that bicycle is a normal choice for this group of people. It is worth mentioning that 40% of long distance commuters (>=25km) spend less than 15 minutes on commuting in 2012, while only 30% of 6-15km commuters can achieve this. It seems that more long distance commuters (>=25km) get shorter commute time than shorter commute distance users (6-15km). We see from Fig. 4 that this phenomenon only happens in City-A. From data in Table 2, the land area in City-A is 82,402.95 $km^2$, but it in City-B is just 11,919.7 $km^2$. The two cities both are Second-tier city. These may suggest that the city's area could influence the speed efficiency of vehicles in a positive way.

People in different distance groups in City-A invest more time on commuting in 2014 than in 2012. For example, the number of people in 0-2km distance group who commute within 15 minutes decreases from 63% in 2012 to 54% in 2014, which indicates the different level of commute time is strong correlated with year. It also demonstrates that the commute time increase with the social development.

From Fig.4, we describe the commute time distribution in every commute distance separately in City-A (2012), City-A (2014), and City-B (2014). Compared the City-A in 2012 with that in 2014, we could get the result that the commute time becomes longer in 2014 than in 2012. Additionally, there is no big difference on commute time distributions of different commute distance groups between morning and night in the same city.

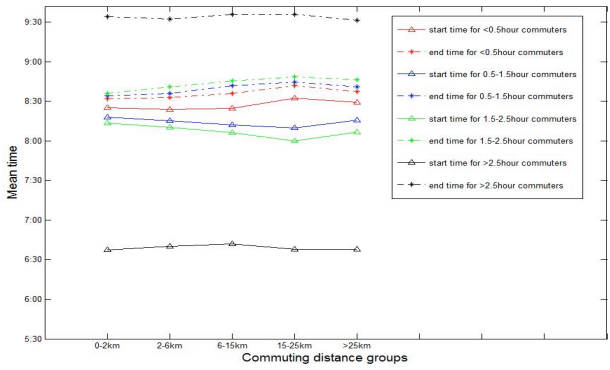

(a) In City-A(2012), commute start time and end time in the morning

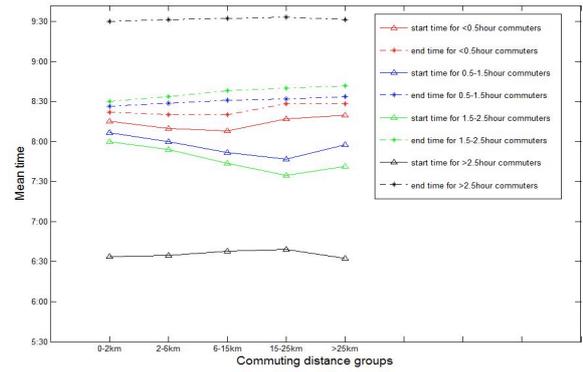

(b) In City-A(2014), commute start time and end time in the morning

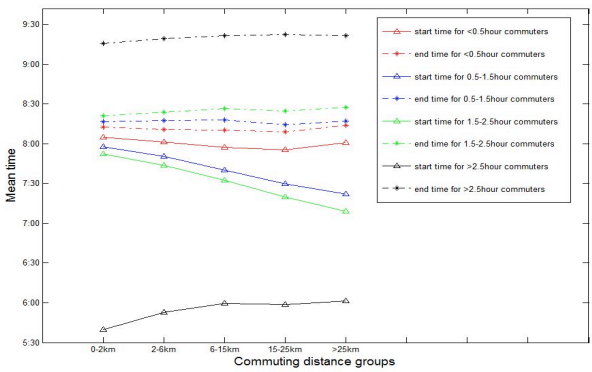

(c) In City-B (2014), commute start time and end time in the morning

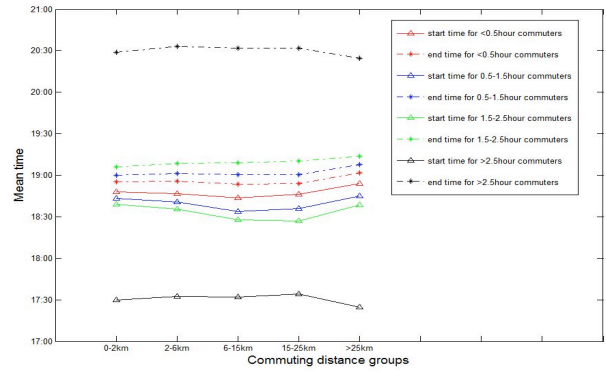

(d) In City-A (2012), commute start time and end time in the evening

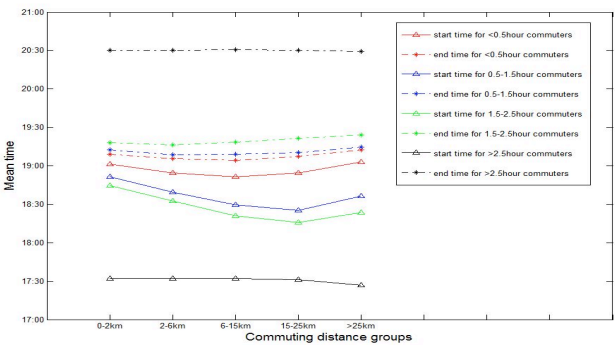

(e) In City-A (2014), commute start time and end time in the evening

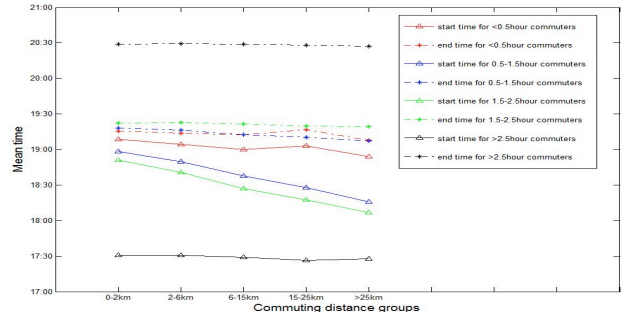

(f) In City-B (2014), commute start time and end time in the evening

Figure 5. Mean commute time period in different commute distance

Fig. 5 illustrates that the mean commute time period regarding different distance groups is in different time ranges, in which a certain color denotes a group of commuters with a certain commute time range. The solid line represents the start commute time, which is recognized by the last record at home in the morning commuting or the last record at workplace at night commuting; the dotted line represents the end commute time, which is recognized by the first record at workplace in the morning commuting or the first record at home at night commuting. The numerical difference between the solid and dotted line is the mean commute time.

From Fig. 5, generally speaking, users who have longer commute distance will start commuting earlier. For users who spend similar commute time, their time to leave home in the morning is earlier in 2014 than two years ago. For example, for less 0.5 hour commute time users, the users in City-A go to work in the morning at 8:20 in 2012, but in 2014 they leave home at 8:10. In City-B, this start time is even approaching to 8:00. Moreover, people's average time to go home are delayed. For less 0.5 hour commute time users in City-A, average time to go home is 18:50 in 2012, and 19:10 in 2014. Moreover, the time to go off work is latest in City-B that is almost approaching19:15. In another words, people in City-B go out earlier, and come home later than in City-A. Actually, City-A is awarded one of ten comfortable cities in China. This result to prove that the peoples live in the City-A have more leisure time than City-B.

**Discussions**

Reports show that City-A and City-B are two most congested cities of China in 2014, while people's commuting behaviors still obey rules of self-adaption.

From the materials in Table 2, we try to analyze the reasons of the difference of people's travel budgets in two cities. Firstly, travel budget seems having no direct quantity relations with population density. Secondly, the transportation conditions like rail traffic mileage and car ownership seem to give explanations. In 2014, the length of rail traffic mileage in City-A is 200km and only 137km in City-B, and the total amount of private car ownership in City-A is 148 million, and 253 million in City-B(data in 2013). While the land area of City-B is nearly one-seventh (11,919.7/82,402.95) of City-A. Worse conditions of railway construction and ground transportation, as well as the more limited city area could be the reason why people in City-B spend more time on commuting. Finally, the increase of GDP is also consistent with the increase of travel budget.

Since mobile traffic usage is ubiquitous, our datasets present its superiority in many aspects against other data collection methods, while there still remains some problems. For example, if there are available wireless access points in people's work places or at home, they may choose to connect to WiFi instead of telecommunication networks, which would result in less traffic records and the trajectories extracted through our method. That could fail to accurately depict human location. Besides, the location denoted by cell towers is not always able to correctly position individuals, since the space-sparsity is still a problem, and even if the dense of cell towers are nearly enough to accurately indicate the users' location at a certain timestamp. Other confounding factors include the underestimation of home/work distance brought by directly using the great circle distance without attaching a correction factor [19]. While, what we have presented is still a novel discovery about the commute patterns of modern cities in China.

Overall, we develop a methodology that allows us to interrogate human commute behaviors. We consider the mobile traffic data as a novel data source for the human mobility behavior research. Reversely, the commute research is also a novel application of mobile traffic data for tracing people's movements. Our future studies should continue focusing on expanding the scope of the datasets to more cities and districts, improving this observation based on more accurate measurements. The universal observations on human mobility in this paper do not only help us gain insights into the fundamental characteristics of how people budget their time, but also have profound policy-level implications about urban and transportation planning.

**Methods**

Firstly, we filter out the effective users through particular regulations, refine their travel trajectories, and investigate the cell tower distributions in these two cities. We estimate the home and work locations through our methodology. Then the commute distance between the home location and work location is found out. Finally, following the Marchetti's constant [14], we research the rule of commute time in different cities between two years to find the underlying invariants factors and the variants factors.

**Dataset**

When the mobile phone users use Internet traffics, we record the items every time, including User ID, the accessing location, start time, end time, etc. In this paper, we choose 3 datasets of mobile traffic records coming from two cities in China. The first dataset is the mobile traffic records from 17th to 23th Nov., 2012 in Chongqin (City-A), covering almost 1.54 million mobile phone users, which is 5.2% of the city's 29.45 million population. The second dataset is the mobile traffic records from 7th to 20th July, 2014 in City-A, covering about 2.2 million mobile phone users, 7.4% of the city's 29.9 million population. And the third one is the mobile traffic records from 5th May to 11th May, 2014 in Tianjin (City-B), covering about 1.58 million mobile phone users, 11.3% of the city's 15.35 million population.

**The cell tower distributions**

By attentively investigating the cell tower distributions in two cities, we find that it exhibits obvious long tail effects, about a half of the cell towers are located very close to their adjacent ones( within 500m), and a wide range of comparatively distant cell towers also contribute to a significant proportion of the total cell towers.

In the Fig. 6, 30% of the Base Stations (BSs) in City-A and 35% of BSs in City-B are within 250m of their adjacent stations, 50% cell towers are within 500m. Only 4% of the cell towers in City-B have adjacent cell towers within 5km to 10km.

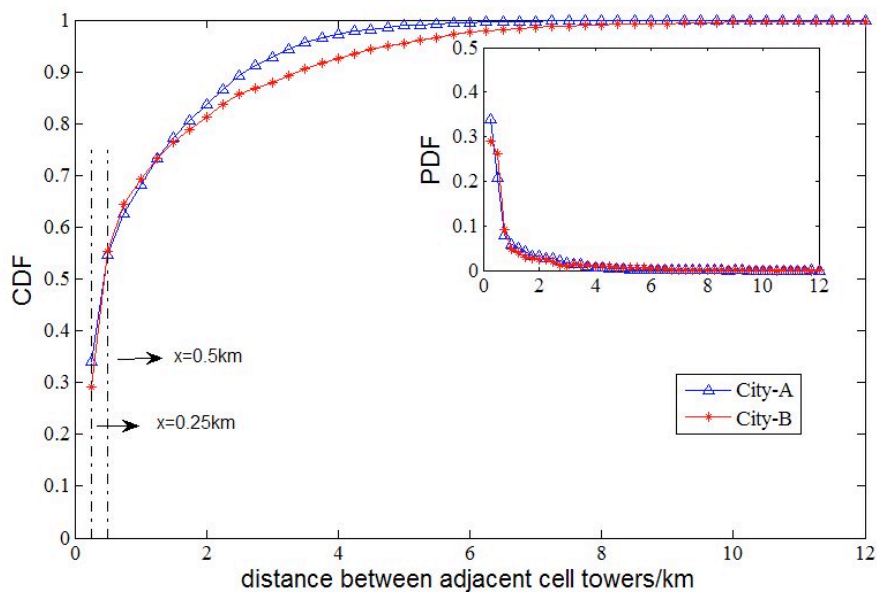

Fig.6 The distribution of the cell towers

The distribution of the BSs in the two cities will have an influence on the spatial accuracy of human trajectories. Too sparse cell distribution would result in terrible space-sparsity effects, i.e., compared with the granularity of their commute distance, the miles between the BS location and the real geographical position of people should not be ignored, especially for short-distance commuters. Based on this consideration, too large inter-cell space would cause upper-estimates of commute distance for users crossing BS and under-estimates for users who do not crossing cell towers. So, people who frequent far-away cell towers, which is not within a radius of 15km of any other cell towers, are filtered out in our method.

Besides, the BSs' distribution gives us a reference on dividing users into different commute distance groups. In order to balance the hierarchy of the separation, cell tower density and the span of the city, we divide people in 0-2km, 2-6km, 6-15km, 15-25km and >25km groups. The people in the first two groups and the middle two groups are respectively regarded as short and medium distance commuters, and people in the last group are long distance commuters.

**Users filtering**

First, users, who don't use traffics more than 1500 records in one day, are filtered out. This guarantees at least 60 records per hour on average. After filtering, there are 8.73% of users left in City-A (2012), 33% of users in City-A (2014) and 21% in City-B (2014).

In order to find human location of home and work, we need to firstly depict human trajectories during the day. People's Trajectories are made up by the locations where mobile traffics are routed. Our analysis is based on the hypothesis that a user stays in a place and does not change his location until generating another record in a different cell, i.e., if an user generates several records in the same location, we suppose that he stays in the original place, and if he generates a new record in a different location, he is regarded to have moved to a new place. Then, based on this perception, users' records can be transferred into trajectories. Note that trajectory starts at the place producing the first record of the day, because we don't know where the user is before any traffic is generated.

Then, we divide the day time into the following three time periods, and find dominant dwelling places as home or work location:

①0:00-7:00, used to find dominant place as the home location in the morning.

②9:00-18:00, used to find dominant place as the work location during the day.

③19:00-24:00, used to find dominant place as the home location at night.

We derived these time periods experimentally and observed that moderate changes to the boundaries do not affect our results. To simplify the calculation, we use ①and② to calculate morning commute time, ②and ③ to calculate night commute time.

In these three time periods, we define dominant location as the location which user stays at least half time of the corresponding time period. That is to say, the location which user spend at least 6 hours during 19:00-0:00-7:00 is regarded as the home location in the morning; the location which user spend at least 4.5hours during 9:00-18:00 is regarded as the work location. Otherwise, we delete users with no dominant home or work locations.

From the following process, people do not have dominant dwelling locations in these time periods are filtering out.

The number of total users in City-A in 2012 is about 1.54 million, increasing to 2.2 million in 2014, and in City-B, the total user number is 1.58 million. Finally, 1.5 % of users in City-A (both in 2012 and 2014) and 3.5 % in City-B are effective samples.

**The computing of commute time**

The period of 6:00-10:00 is defined as the morning commuting period, and 17:00-21:00 as night commuting period. We define commuting activity happens between 6:00-10:00 and 17:00-21:00, so we check the last traffic record at home and the first traffic record in work place within 6:00-10:00, the duration between the time stamp of this two record is defined as the estimation of the user's morning commute time. Vice versa, the duration between the last traffic record in work place and the first traffic record at home within 17:00-21:00 is defined as the user's night commute time. After this process, Users without records during commute periods in pre-identified home or work locations are abandoned.

Our analysis is based on the hypothesis that people leave the home or work place the minute after generating a record, and they generate a record right upon arriving at work or home place. As a result, this method only gives an upper estimation of the actual commute time.

**Acknowledgments**

This work was supported by the National Natural Science Foundation of China (61201153), the National 973 Program of China under Grant (2012CB315805), the Prospective Research Project on Future Networks in Jiangsu Future Networks Innovation Institute (BY2013095-2-16), the National Basic Research Program 973 of China (2012CB315801), and the Fundamental Research Funds for the Central Universities (2013RC0113).

**Tables**

Table 1.The proportions of people in different distance groups. The listed percentage in the Table is the corresponding proportion in three groups with all the valid user.

| The final effective users in different distance groups | | | |
|---|---|---|---|
| | City-A(2012) | City-A(2014) | City-B(2014) |
| 0-2km | 4,800(56.5%) | 20,000(51%) | 11,000(53%) |
| 2-6km | 2,000(23.5%) | 9,800(25%) | 5,000(24.5%) |
| 6-15km | 1,200(14%) | 6,700(17%) | 3,000(14.5%) |
| 15-25km | 250(3%) | 1,550(4%) | 800(4%) |
| >=25km | 250(3%) | 1,100(3%) | 800(4%) |
| Sum effective Users | 8,500 (0.55%) | 39,000 (1.8%) | 20,600 (1.3%) |
| Total users in records | 1,540,000 | 2,200,000 | 1,580,000 |

Table 2. Characteristics of the two Cities

| City and year | City-A 2012 | City-B 2012 | City-A 2014 | City-B 2014 |
|---|---|---|---|---|
| Population | 29,450,000 | 14,130,000 | 29,900,000 | 15,350,000 |
| Number of users( The ratio in population | 1,540,000(5.2%) | 1,190,000 (in 2007) | 2,200,000(7.4%) | 1,580,000(11.3%) |
| Area of land | 82,402.95 km2 | 11,919.70 km2 | 82,402.95 km2 | 11,919.70 km2 |

| Population density | 357.39 persons/km2 | 1185.43 persons/km2 | 362.85persons/km2 | 1287.78persons/km2 |
|---|---|---|---|---|
| GDP | 1,145,900million | 1,288,518million | 1,400,000million | 1,560,300million |
| Private car ownership | 117 million | 236million | In 2013<br>148 million | In 2013<br>253 million |
| Rail traffic mileage | 131km | 94km | 200km | 137km |
| Traffic congestion situation | Normal | serious<br>( 8 serious traffic jam points) | serious<br>( 2 serious traffic jam points) | serious |
| Urban division | Second-tier city | Second-tier city | Second-tier city | Second-tier city |

Table 3. Conclusion regarding commute time constant

|  | Meet a constant | | Not Meet a constant | | Marchetti's constant |
|---|---|---|---|---|---|
|  | Commute distance | Commute time of single way (morning-night) | Commute distance | Commute time of single way |  |
| City-A(2012) | >18km subscribers | 0.80−0.84 hour | 0-18km subscribers | Increase with distance | 1.18 hour |
| City-A(2014) | >18km subscribers | 1.34-1.39 hour | 0-18km subscribers | Increase with distance | 1.59 hour |
| City-B(2014) | >18km subscribers | 1.50-1.70 hour | 0-18km subscribers | Increase with distance | 1.74 hour |

# Author information

**Contributions**

Hongyan Cui, Stanislav Sobolevskv and Carlo Ratti focus on method planning and confirmation, analyze results. Yuxiao Wu writes and debugs programs to apply the method. Shuai Xu realizes visualization.

**Competing interests**

The authors declare no competing financial interests.

**Corresponding author**

Correspondence to Hongyan Cui(cuihy@bupt.edu.cn).